\documentclass[12pt,epsf]{article}

\usepackage{epsfig,amssymb}
\usepackage{graphicx,color}

\newcommand{\be}{\begin{equation}}
\newcommand{\ee}{\end{equation}}
\newcommand{\bear}{\begin{eqnarray}}
\newcommand{\eear}{\end{eqnarray}}
\newcommand{\ba}{\begin{array}}
\newcommand{\ea}{\end{array}}
\newcommand{\nn}{\nonumber}

\begin{document}
\begin{titlepage}
\vspace{.5in}
\begin{flushright}
TCDMATH 07--12
\end{flushright}
\vspace{0.5cm}

\begin{center}
{\Large\bf Fluctuations around the Tachyon Vacuum \\
in Open String Field Theory}\\
\vspace{.5in}

{\rm{O-Kab Kwon,}$^{*}$}\footnote{\it email:\,okabkwon@maths.tcd.ie}\,\,
{\rm{Bum-Hoon Lee,}$^{\dag\S}$}\footnote{\it email:\,bhl@sogang.ac.kr}\,\,
{\rm{Chanyong Park,}$^{\S}$}\footnote{\it email:\,cyong21@sogang.ac.kr}\,\,
{\rm{and Sang-Jin Sin}$^{\ddag}$}\footnote{\it email\,:sjsin@hangyang.ac.kr}\\

\vspace{.1in}
{\small * \it School of Mathematics, Trinity College, Dublin, Ireland}\\
{\small \dag \it Department of Physics, Sogang University, 121-742, Seoul,
  Korea}\\
{\small \S \it Center for Quantum Spacetime, Sogang University,
  121-742, Seoul,
  Korea}\\
{\small \ddag \it  Department of Physics, Hanyang University, 133-791,
  Seoul, Korea}\\

\vspace{.5in}
\end{center}
\begin{center}
{\large\bf Abstract}
\end{center}
\begin{center}
\begin{minipage}{4.75in}
We consider quadratic fluctuations around the tachyon vacuum numerically
in open string field theory. We work on a space ${\cal H}_N^{{\rm vac}}$
spanned by basis string states used in the Schnabl's vacuum solution.
We show that the truncated form of the Schnabl's vacuum solution
on ${\cal H}_N^{{\rm vac}}$ is well-behaved in numerical work.
The orthogonal basis for the new BRST operator $\tilde Q$
on ${\cal H}_N^{{\rm vac}}$ and the quadratic forms of potentials
for independent fields around the vacuum are obtained.
Our numerical results support that the Schnabl's vacuum solution
represents the minimum energy solution for arbitrary fluctuations also
in open string field theory.
\end{minipage}
\end{center}
\end{titlepage}

\newpage
\section{Introduction}
After Schnabl's analytic proof for Sen's first
conjecture~\cite{Schnabl:2005gv} in Witten's cubic open string
field theory (OSFT)~\cite{Witten:1985cc}, there has been remarkable
progress in analytic understanding of
OSFT~\cite{Okawa:2006vm}-\cite{Rastelli:2007gg}.
In particular, Sen's third conjecture was proved analytically using
the exactness of identity string state~\cite{Ellwood:2006ba}.
Analytic solutions for marginal deformations, especially,
rolling tachyon solution, were constructed~\cite{Schnabl:2007az}
and extended to superstring field theory~\cite{Erler:2007rh}.
General formalism for the marginal deformations including
the case of singular operator products was
constructed~\cite{Kiermaier:2007vu}.
See also ref.~\cite{Fuchs:2007yy} for other approaches in marginal
deformations. And off-shell Veneziano amplitude in OSFT was
calculated by employing a definition of the open string propagator
in the Schnabl's gauge~\cite{Fuji:2006me,Rastelli:2007gg}.

In this paper, we consider quadratic fluctuations written as
$\tilde Q$-term in the action of OSFT numerically, construct
an orthogonal basis, and investigate the stability of Schnabl's
vacuum solution and the structure of tachyon vacuum.
Here $\tilde Q$ is a new BRST operator defined at the tachyon
vacuum, which is composed of the original BRST operator
$Q_B$ and tachyon vacuum solution $\Psi$.
In virtue of the exact expression of the vacuum string field
given by Schnabl~\cite{Schnabl:2005gv}, we construct the
$\tilde Q$-term for arbitrary fluctuations in a subspace spanned
by wedge state with operator insertions.

Before Schnabl's breakthrough~\cite{Schnabl:2005gv}, there
were many trials to understand the properties of $\tilde Q$
without exact expression of the vacuum
solution~\cite{Rastelli:2000hv}-\cite{Giusto:2003wc}.
Most of works in this area were devoted to the proof of vanishing
cohomology of $\tilde Q$~~\cite{Ellwood:2001py,
Hata:2001rd,Ellwood:2001ig,Giusto:2003wc,Imbimbo:2006tz} regarding
to Sen's third conjecture~\cite{Sen:1999mh,Sen:1999xm}.
As one of recent main analytic progresses of OSFT,
the vanishing cohomology of $\tilde Q$ was proved by showing that all
$\tilde Q$-closed states are $\tilde Q$-exact.
Therefore, all fluctuation fields around the vacuum are off-shell ones
according to the proof.
To study the stability of the Schnabl's vacuum solution and
the tachyon vacuum structure in terms of potentials for
independent fields, we restrict our interest to the spacetime
independent gauge fixed off-shell fluctuations
in $\tilde Q$-term neglecting the cubic interactions for
the fluctuations.

In order to describe the physics around the vacuum completely, we have to
take into account the string fluctuations governed by the $\tilde Q$-term
on the full Hilbert space of OSFT. However, in numerical work,
we have to take a subspace of the full Hilbert space by an appropriate
approximation such as the well-known level truncation
approximation~\cite{Kostelecky:1989nt}-\cite{Gaiotto:2002wy}.
In this work, we consider a truncated subspace spanned by basis
string states which are used in Schnabl's vacuum solution.
Since the every basis state satisfies the Schnabl's gauge
condition, all fluctuations on the basis satisfy
the Schnabl's gauge condition.
The vacuum solution is expressed by an infinite series in terms of
wedge states with operator insertions.
In construction of our truncated subspace, ${\cal H}_N^{{\rm vac}}$,
we truncate the basis states up to wedge state $|N+2\rangle$
with operator insertions.

In section 2, we introduce a truncated Schnabl's solution on
${\cal H}_N^{{\rm vac}}$ to use the Schnabl's solution in numerical work.
In $N\to\infty$ limit, the truncated Schnabl's solution becomes
the exact one. We examine the convergence and accuracy of the truncated
Schnabl's solution in BPZ inner product by increasing $N$.

In section 3, we consider spacetime independent arbitrary quadratic
fluctuations and obtain orthogonal basis of $\tilde Q$ using
the symmetric property of $\tilde Q$ on ${\cal H}_N^{{\rm vac}}$.
We investigate the numerical properties of $\tilde Q$ for various
situations, discuss the stability of Schnabl's vacuum solution, and
find quadratic forms of potential for
independent fields around the tachyon vacuum.
We conclude in section 4.

\section{Truncated Schnabl's Solution}

We begin with a brief review of OSFT and an introduction of
Schnabl's analytic vacuum solution. The action of
OSFT~\cite{Witten:1985cc} has the form
\bear\label{AC}
S(\Phi) = - \frac1{g_o^2} \left[\frac12 \left<\Phi,\,Q_{{\rm B}}
\Phi\right> + \frac13\left<\Phi,\,\Phi*\Phi\right>\right],
\eear
where $g_o$ is the open string coupling constant, $Q_{{\rm B}}$ is the
BRST operator, `$*$' denotes Witten's star product, and
$\langle\cdot,\cdot\rangle$ is the BPZ inner product.
In this definition of BPZ inner product, we omit the spacetime
volume factor. The action (\ref{AC}) is invariant
under the gauge transformation
$\delta \Phi = Q_{{\rm B}}\Lambda + \Phi*\Lambda -\Lambda*\Phi$
for any Grassmann-even ghost number zero state $\Lambda$ and satisfies
the classical field equation,
\bear\label{EOM}
Q_{{\rm B}}\Phi + \Phi*\Phi = 0.
\eear
Schnabl's analytic vacuum solution
of the Eq.~(\ref{EOM}) was represented as~\cite{Schnabl:2005gv}
\bear\label{SS}
\Psi\equiv \lim_{N\to\infty}\left[\sum^N_{n=0}
\psi_n^{'}-\psi_N\right],
\eear
where $\psi_n $ and $\psi_n^{'}\equiv \frac{\partial\psi_n}{\partial n}$
are the wedge state
$|n+2\rangle$~\cite{Rastelli:2000iu,Rastelli:2001vb}
with operator insertions, given by
\bear\label{gss}
&& \psi_0 = \frac{2}{\pi} c_1 |0\rangle,
\nn \\
&& \psi_n = \frac{2}{\pi} c_1 |0\rangle *
|n\rangle * B_1^L c_1 |0\rangle,
\qquad (n\ge 1)
\nn \\
&&\psi_0^{'}= K_1^L c_1 |0\rangle + B_1^L c_0c_1|0\rangle,
\nn \\
&&\psi_n^{'}=c_1|0\rangle * K_1^L |n\rangle*B_1^Lc_1 |0\rangle,
\qquad (n\ge 1).
\eear
Here we use the following operator representations on
upper half plane(UHP),
\bear
&&B_1^L =\int_{C_L}\frac{d\xi}{2\pi i}(1+\xi^2) b(\xi),
\nn \\
&&K_1^L =\int_{C_L}\frac{d\xi}{2\pi i}(1+\xi^2) T(\xi),
\eear
where $b(\xi)$ is the $b$ ghost and the contour $C_L$ runs
counterclockwise along the unit circle with ${\rm Re}\, z <0$.
In obtaining the solution $\Psi$, Schnabl used clever
coordinate $z=\tan^{-1}\xi$ and gauge choice
\bear\label{SG}
{\cal B}_0\Psi=0,
\eear
where ${\cal B}_0 = \oint\frac{d\xi}{2\pi i}
(1 + \xi^2)\tan^{-1}\xi \,b(\xi)$.

We can describe the physics around the tachyon vacuum $\Psi$
by shifting the string field $\Phi=\Psi + \tilde \Psi$.
Then the action in terms of string field $\tilde\Psi$ is given by
\bear \label{AC2}
\tilde S(\tilde\Psi)\equiv S(\Psi + \tilde\Psi) - S(\Psi)
=- \frac12 \langle\tilde\Psi,\,\tilde Q\tilde \Psi\rangle
- \frac13\langle\tilde\Psi,\,\tilde\Psi*\tilde\Psi\rangle,
\eear
where we set $g_o=1$ for simplicity.
The new BRST operator $\tilde Q$ acts on a string field $\phi$ of
ghost number $n$ through
\bear\label{tilQ}
\tilde Q \phi = Q_{{\rm B}}\phi + \Psi*\phi
-(-1)^n \phi*\Psi.
\eear
It is straightforward to check the nilpotent property of $\tilde Q$
using the properties of the star products and
the equation of motion for $\Psi$, $Q_{{\rm B}}\Psi
+ \Psi*\Psi=0$. The new action for the string field
$\tilde \Psi$ has the same form as the original action (\ref{AC}) when
$Q_{{\rm B}}$ and $\Psi$ are replaced by $\tilde Q$ and $\tilde \Psi$
respectively.
So we can easily find the form of the gauge transformation for the action,
$\delta\tilde\Psi = \tilde Q\tilde\Lambda + \tilde\Psi*\tilde\Lambda
-\tilde\Lambda*\tilde\Psi$, with any Grassmann-even ghost number zero
state $\tilde\Lambda$.

Our purpose in this paper is to investigate the physical properties of the
new action $\tilde S(\tilde\Psi)$ around the tachyon vacuum neglecting
the cubic term in Eq.~(\ref{AC2}) numerically.
To accomplish this purpose, we have to use the Schnabl's
solution according to the definition of $\tilde Q$ in Eq.~(\ref{tilQ}).
Most difficulties in numerical computations by using Schnabl's solution
come from the infinite series expression of it given in Eq.~(\ref{SS}).
To use the Schnabl's solution in numerical work we have to truncate
the infinite series somehow. As a truncation approximation
similar to the well-known level truncation approximation
in open string field
theory~\cite{Kostelecky:1989nt,Sen:1999nx, Moeller:2000xv, Gaiotto:2002wy},
we consider the following wedge state truncation
for the solution (\ref{SS}),
\bear\label{tcSS}
\Psi_{N} =  \sum_{n=0}^{N} \psi_n^{'}-\psi_{N},
\eear
where $N$ is a finite number.\footnote{In the level truncated
OSFT~\cite{Kostelecky:1989nt}-\cite{Gaiotto:2002wy},
the open string fields are restricted to modes with $L_0$ eigenvalues
which are smaller than the maximum level $L$. Thus the resulting solutions
for various numbers of $L$ have different forms. But in our case we
truncate the known exact solution without change of coefficients for
basis states.}
We include the string states up to wedge state $|N+2\rangle$
in the truncated Schnabl's solution (\ref{tcSS}).
In this representation, the Schnabl's solution $\Psi$ corresponds
to $\Psi_\infty$.
%%%%%%%%%%%%%%%%%%%%%%%%%%%%%%%%%%%%%%%%%%%%%%%%%%%%%%%%%%%%%
\begin{table}\label{atSS}
\begin{center}\def\st{\vrule height 3ex width 0ex}
\begin{tabular}{|c|c|c|c|c|c|c|} \hline

$N$ & $0$ & $2$ & $4$ &
$6$ & $8$ & $10$
\st\\[0ex] \hline

$f(N)$&$-0.11289$&$-0.73227$ & $-0.89030$ &
$-0.94163$ & $-0.96400$ & $-0.97564$
\st \\ \hline \hline

$N$ & $20$ & $40$ & $60$ &
$80$ & $100$ & $200$
\st\\[0ex] \hline

$f(N)$&$-0.99319$ & $-0.99820$ & $-0.99919$ & $-0.99954$ & $-0.99970$
& $-0.99992$
\st\\[0ex] \hline
\end{tabular}
\end{center}
\caption{{\small Values of $f(N)$ for various truncation numbers.}}
\end{table}
%%%%%%%%%%%%%%%%%%%%%%%%%%%%%%%%%%%%%%%%%%%%%%%

To use the truncated Schnabl's solution instead of the tachyon vacuum
solution $\Psi$ given in Eq.~(\ref{SS}) in numerical computations,
we have to check the properties of $\Psi_N$ in BPZ inner products.
We insert $\Phi=\Psi_N$ into the action (\ref{AC}), and increase $N$
to figure out the properties of $\Psi_N$ in BPZ inner products.
We compare this result with the well-known explicit result by
Schnabl~\cite{Schnabl:2005gv}. It was proved that $\Psi$
in Eq.~(\ref{SS}) reproduces the exact tension ($=1/2\pi^2$
in $\alpha'=1$ unit) of D25-brane expected by Sen's first conjecture,
i.e.,
\bear\label{AC3}
S(\Psi) = -\frac12
\left<\Psi,\,Q_{{\rm B}}\Psi\right>
- \frac13\left<\Psi,\,\Psi*\Psi\right> =
\frac1{2\pi^2},
\eear
where
\bear\label{SQBS}
&&\langle\Psi,\,Q_{{\rm B}}\Psi\rangle =
-\frac{3}{\pi^2}, \quad
\langle\Psi,\,\Psi*\Psi\rangle= \frac{3}{\pi^2}.
\eear

In Table 1, we give the values of the normalized tachyon
potential~\cite{Sen:1999xm}, $f(N)$, defined as
\bear\label{ntp}
f(N)\equiv - 2\pi^2 S(\Psi_N)
\eear
for various truncation numbers.
From this numerical result, we see that the quantities $f(N)$
converge to the exact value $f(\infty)=-1$ with high accuracy
as we increase $N$.
%%%%%%%%%%%%%%%%%%%%%%%%%%%%%%%%%%%%%%%%%%%%%%%%%%%%%%%%%%%%%%%%
\begin{figure}\label{Fig1}
\centerline{\includegraphics[width=115mm]{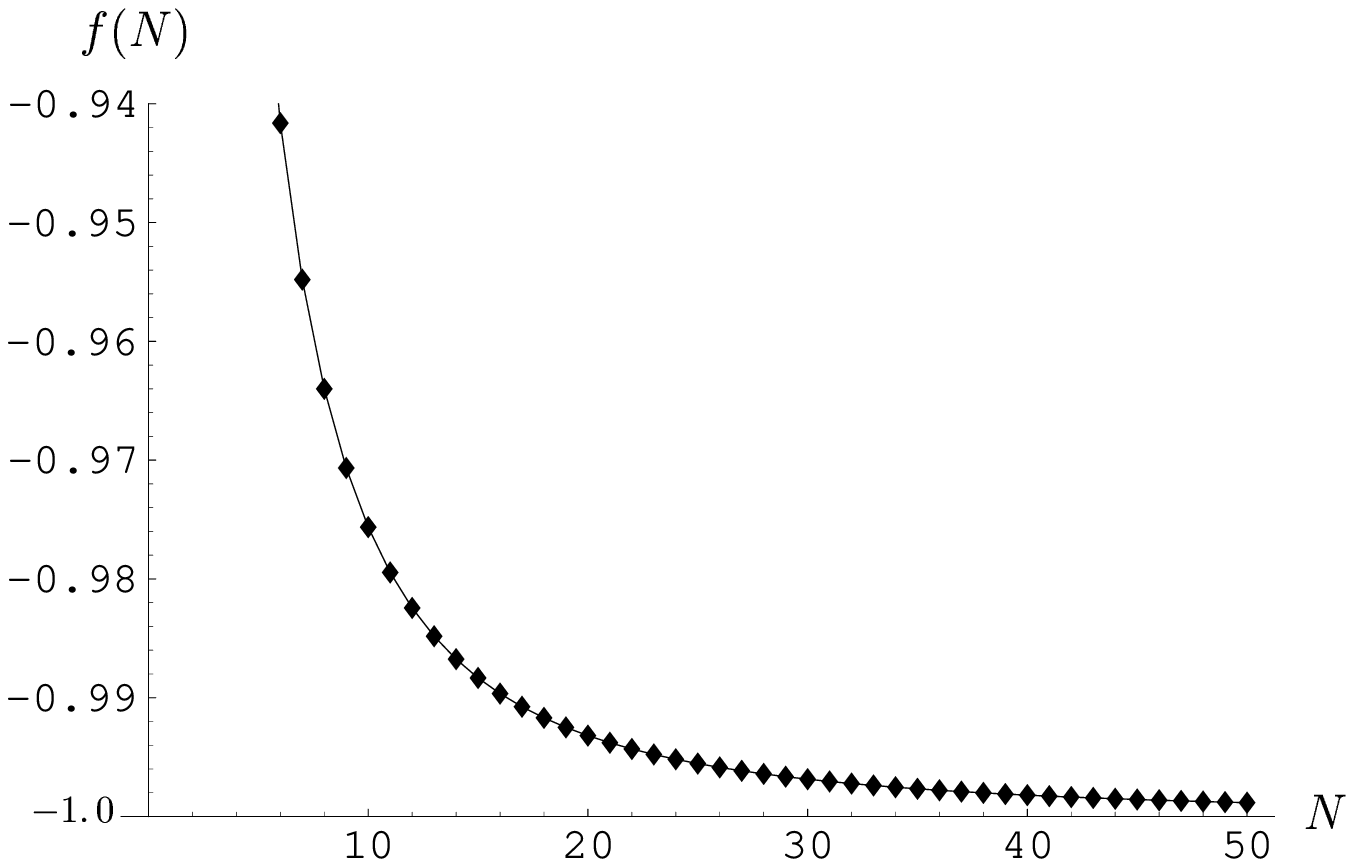}}
\caption{{\small Graph of $f(N)$. The points represent $f(6),\, f(8),\,
\cdots f(50)$ from the left. }}
\end{figure}
%%%%%%%%%%%%%%%%%%%%%%%%%%%%%%%%%%%%%%%%%%%%%%%%%%%%%%%%%%%%%%%%
In the usual level truncation approximation in OSFT, the normalized
tachyon potential approaches to $-1$ as $L\to\infty$
non-monotonically~\cite{Taylor:2002fy,Gaiotto:2002wy}.
But in this wedge state truncation, $f(N)$ is a monotonic function
with respective to $N$. In Fig.1, we plot the behavior of $f(N)$.
Therefore, we can safely replace the infinite series of
Schnabl's solution $\Psi$ with the truncated Schnabl's solution $\Psi_N$
for sufficiently large number of $N$ in the numerical computations
of the BPZ inner products which contain the Schnabl's solution.

\section{Quadratic Fluctuations around the Tachyon Vacuum}\label{num}

Small fluctuations of string field $\tilde\Psi$ around the
tachyon vacuum are governed by quadratic term in the action (\ref{AC2}),
\bear\label{AC4}
\tilde S_0(\tilde \Psi) =
- \frac12 \langle\tilde\Psi,\,\tilde Q\tilde \Psi\rangle.
\eear
This action is composed of innumerable fields which are related
each other in general.
In this section, we investigate the properties of the spacetime
independent fluctuations of $\tilde \Psi$.
To do this we calculate the quantity
$\langle\tilde\Psi,\,\tilde Q\tilde\Psi\rangle$ numerically.
In this calculation, we restrict our interests to arbitrary
gauge fixed fluctuations with ghost number 1
on the space spanned by wedge states with some operator insertions,
$\psi_m^{'}$, $(m=0,1,2,\cdots)$, used in the expression
of Schnabl's solution (\ref{SS}).
We construct the orthogonal basis of $\tilde Q$, which allows
to define independent fields and
obtain the quadratic potentials of the fields.

\subsection{Orthogonal Basis of $\tilde \Psi$}

In principle we have to consider the fluctuation field
$\tilde\Psi$ on the full Hilbert space around the
tachyon vacuum
to study the physical properties of $\tilde \Psi$ given
in the action (\ref{AC4}).
However, the Hilbert space around the vacuum
is not well-known up to now.
In our numerical work we restrict our interests
to the fluctuation field $\tilde \Psi$ on the subspace spanned
by basis states,
\bear\label{BS1}
{\cal H}^{{\rm vac}}_{N} \equiv {\rm span}\{\psi_n^{'},\,
0\le n \le N\}
\eear
with large but finite number of $N$.

Actually we can express the ordinary piece
$\sum_{n=0}^\infty \psi_n^{'}$ on the subspace
${\cal H}_\infty^{{\rm vac}}$ without the phantom piece $-\psi_\infty$
in Schnabl's solution (\ref{SS}).
The ordinary piece alone contributes to the vacuum energy about
$50\%$ and the remaining contributions come from the interactions
between the ordinary piece and phantom one and the phantom piece alone.
So, the contribution of the phantom piece to the vacuum energy is
nontrivial~\cite{Schnabl:2005gv,Okawa:2006vm}.
If we parametrize the phantom piece like as $\alpha\psi_\infty$,
we can easily see that the vacuum energy has the minimum value for the
case $\alpha=-1$ which is the exact coefficient of $\psi_\infty$ in
the phantom piece. Thus the energy contributions of the fluctuations
along the direction of the phantom piece is positive always.
However, the fluctuations which are expressed by the basis on
${\cal H}_\infty^{{\rm vac}}$ and $\psi_\infty$ simultaneously
have nontrivial difficulties in the investigation of energy contributions
by using our method which will be explained later.
In this reason, we restrict our interests in the fluctuations on
${\cal H}_N^{{\rm vac}}$ only. We believe that this subspace is very
important space in the tachyon condensation as we explained.

We express $\tilde\Psi$ on the truncated subspace spanned by
$(N+1)$-basis states (\ref{BS1})
as
\bear\label{tpsi}
\tilde \Psi_N =  \sum_{n=0}^{N} c_n\psi_n^{'},
\eear
where $c_n$ is an arbitrary small real number.
Since each basis state satisfies
\bear\label{SG2}
{\cal B}_0\,\psi_n^{'}=0, \quad (n\ge 0),
\eear
the fluctuation $\tilde \Psi_N$
in Eq.~(\ref{tpsi}) satisfies the gauge choice
\bear\label{SG3}
{\cal B}_0\tilde\Psi_N=0.
\eear
In other words, we consider the gauge fixed fluctuations
around the tachyon vacuum.

Inserting the Eq.~(\ref{tpsi}) into the quantity
$\langle\tilde\Psi,\,\tilde Q\tilde\Psi\rangle$ in Eq.~(\ref{AC4}),
we obtain
\bear\label{PQP1}
\langle\tilde\Psi,\,\tilde Q\tilde\Psi\rangle_N
= \sum_{m=0}^{N}\sum_{n=0}^{N} c_m c_n (\tilde Q_N)_{mn},
\eear
where
\bear\label{Qmn}
(\tilde Q_N)_{mn} &\equiv& \langle\psi_m^{'},\,
\tilde Q_N\psi_n^{'}\rangle
\nn \\
&=& \langle\psi_m^{'},\, Q_{{\rm B}}\psi_n^{'}\rangle
+  \langle\psi_m^{'},\,\Psi_N * \psi_n^{'}\rangle
+ \langle\psi_m^{'},\,  \psi_n^{'}*\Psi_N\rangle
\nn \\
&=& \frac{\partial}{\partial m}\frac{\partial}{\partial n}\, f(m,n)
+ 2 \sum_{k=0}^{N} \frac{\partial}{\partial m}
\frac{\partial}{\partial k}\frac{\partial}{\partial n}\,h(m,k,n)
-2 \frac{\partial}{\partial m}\frac{\partial}{\partial n}\,h(m,N,n)
\eear
with $\Psi_N$ given in Eq.~(\ref{tcSS}).
Here $f(m,n)$ and $h(m,k,n)$ are the explicitly known
formulae~\cite{Schnabl:2005gv,Okawa:2006vm},
\bear
f(m,n)&\equiv& \langle\psi_n,Q_{{\rm B}}\psi_m\rangle
\nn \\
&=&  \frac{1}{\pi^2} \,
\left( 1 + \cos \frac{(m-n)\pi}{m+n+2} \right)
\left( -1 + \frac{m+n+2}{\pi} \sin \frac{2 \, \pi}{m+n+2} \right)
\nonumber \\
&& {}+ 2 \sin^2 \frac{\pi}{m+n+2}
\left[ \, -\frac{m+n+1}{\pi^2}
+ \frac{m n}{\pi^2} \cos \frac{(m-n)\pi}{m+n+2}\right.
\nn \\
&& \hskip 3.8cm \left. + \frac{(m+n+2)(m-n)}{2 \, \pi^3} \sin
\frac{\pi \, (m-n)}{m+n+2} \,
\right] \,,
\label{pQBp} \\
h(m,k,n) &\equiv&\langle\psi_n,\psi_m*\psi_k\rangle
\nn \\
&=& \frac12\left(\frac2{\pi}\right)^7
(m+n+k+3)^2\sin^2\left(\frac{\pi}{m+n+k+3}\right)
\nn \\
&&\times\sin\left(\frac{(n+1)\pi}{m+n+k+3}\right)
\sin\left(\frac{(m+1)\pi}{m+n+k+3}\right)
\sin\left(\frac{(k+1)\pi}{m+n+k+3}\right).
\label{ppp}
\eear
In the last step of Eq.(\ref{Qmn}) we used the symmetries among indices
$m$, $k$, and $n$ in $h(m,k,n)$.
which come from the twist symmetry of OSFT.
Using the properties of the
BPZ inner product and BRST operator $Q_B$, we can see that there is a
symmetry between $m$ and $n$ in $f(m,n)$ also.
So $(\tilde Q_N)_{nm}$ is a matrix element of
the real symmetric $(N +1) \times (N +1)$ matrix $\tilde Q_N$.

Since $\tilde Q_N$ is a finite dimensional
real symmetric matrix, we can diagonalize $\tilde Q_N$ according to
the following finite dimensional spectral theorem:

\noindent
{ \it To every finite dimensional real symmetric matrix $A$ there exists a
real orthogonal matrix $\tilde U$ such that $D= \tilde U A \tilde U^{T}$
is a diagonal matrix}.

\noindent
Here $\tilde U^T= \tilde U^{-1}$ is the transpose matrix of $\tilde U$.
According to this theorem, we can diagonalize the matrix $\tilde Q_N$
by an orthogonal matrix $U$ as
\bear\label{DQ1}
\tilde Q_N=U^T\tilde Q_N^{(d)} U,
\eear
where $\tilde Q_N^{(d)}$ is a diagonalized matrix.
Substituting the relation (\ref{DQ1}) into (\ref{PQP1}),
we obtain
\bear\label{PQP3}
\langle\tilde\Psi,\,\tilde Q\tilde\Psi\rangle_N &=&
\sum_{m=0}^N\sum_{n=0}^N c_m
( U^T \tilde Q_N^{(d)} U)_{mn} c_n
\nn \\
&=& \sum_{m=0}^N\bar\lambda_m\bar c_m^2 ,
\eear
where the values $\bar\lambda_m$ are diagonal components of
$\tilde Q_N^{(d)}$, i.e., eigenvalues of $\tilde Q_N$, and we define
\bear\label{TC}
&&\bar c_m = \sum_{n=0}^N U_{mn}c_n.
\eear
Since $c_m$ and all matrix elements $U_{mn}$ are real,
the arbitrary coefficients $\bar c_m$ are also real.
By comparing (\ref{PQP1}) with (\ref{PQP3}) and using the
property of orthogonal matrix, $U^{T} = U^{-1}$, we obtain
\bear\label{obs2}
\langle\bar\psi_m,\,\tilde Q\bar\psi_n\rangle_N
=\bar\lambda_m \delta_{mn},
\eear
where the orthogonal basis $\bar\psi_m$ is defined as
\bear\label{obs1}
\bar\psi_m = \sum_{n=0}^N U_{mn}\psi_n^{'}.
\eear
In the orthogonal basis (\ref{obs1}), the truncated
Schnabl's solution (\ref{tcSS}) and the fluctuation
string field (\ref{tpsi}) are written respectively as,
\bear\label{tcSS2}
\Psi_N=\sum_{m=0}^N\sum_{n=0}^N U_{nm}\bar\psi_n - \psi_N,
\qquad \tilde\Psi=\sum_{m}^N\bar c_m\bar\psi_m.
\eear

\subsection{Numerical results}

To determine the diagonal components $\bar\lambda_m$ and the
orthogonal matrix $U$ for a given truncation number $N$, we calculate
the matrix elements of $\tilde Q_N$ given in Eq.~(\ref{Qmn})
with the assistance of the MATHEMATICA program.
During all processes of the numerical computations, we adjusted the
number of significant digits by manipulating options of
the program to increase numerical precisions.
We calculate the matrix components of $\tilde Q_N$ up to
$N=200$. In principle, we can obtain numerical results
for the higher number of $N$ than $N=200$.
As we will see in the numerical data, however, we can capture most
of characteristic features of $\tilde Q_N$ by using the data
up to $N=200$ sufficiently.

In our setup, we truncate the infinite dimensional matrix
representation of $\tilde Q$ to a finite dimensional $(N+1)\times
(N+1)$ matrix $\tilde Q_N$ with truncated Schnabl's solution
(\ref{tcSS}) for numerical work.
Since the vacuum energy calculated from the truncated Schnabl's
solution $\Psi_N$ converges to the exact one by raising $N$
without any
singularities~\cite{Schnabl:2005gv,Okawa:2006vm},
the convergence of a certain quantity can be a criterion whether
it is meaningful or not.

From the numerical results of $\tilde Q_N$,
we determine the eigenvalues $\bar\lambda_m$
 and the orthogonal matrix $U$
for a given $N$. In Table 2, we give eigenvalues of $\tilde Q_N$
for several truncation numbers of $N$.\footnote{$\bar\lambda_0$,
$\bar\lambda_1$, $\bar\lambda_2$, $\cdots$ is a descending series
for the positive eigenvalues of $\tilde Q_N$. The negative
eigenvalue is named as $\bar\lambda_-$.}
For positive eigenvalues, all eigenvalues, $\bar\lambda_0$,
$\bar\lambda_1$, $\bar\lambda_2$, $\cdots$, seem to converge rapidly,
i.e., $\bar\lambda_m$ for a given $m$ has
a convergent series by raising $N$. For example,
we explicitly show the convergent properties of $\bar\lambda_m$
for several largest eigenvalues in Table 3.
We can also see the convergency of $\bar\lambda_m$
for a given $m$ graphically in Fig.~2.

Similarly to the eigenvalues of $\tilde Q_N$, the expansion
coefficients of the orthogonal basis $\bar\psi_m$ with
positive eigenvalues in Eq.~(\ref{obs1})
have convergent series as we increase $N$.
For example, we give the lowest orthogonal state $\bar\psi_0$
which gives the largest eigenvalues $\bar\lambda_0$ for various
truncation numbers of $N$,
\bear\label{NN}
N=0:&& \bar\psi_0 = \psi_0^{'},
\nn \\
N=1:&& \bar\psi_0 = 0.8123\psi_0^{'} + 0.5832\psi_1^{'},
\nn \\
N=2:&& \bar\psi_0 =0.9309\psi_0^{'} + 0.2798\psi_1^{'}
-0.2349\psi_2^{'},
\nn \\
N=3:&& \bar\psi_0 =0.8753\psi_0^{'} + 0.1228\psi_1^{'}
-0.3239\psi_2^{'} - 0.3374\psi_3^{'},
\nn \\
N=4:&& \bar\psi_0 =0.8521\psi_0^{'} + 0.1912\psi_1^{'}
-0.2686\psi_2^{'}-0.3188\psi_3^{'}-0.2521\psi_4^{'}
\nn \\
N=5:&& \bar\psi_0 =0.8384\psi_0^{'} + 0.2503\psi_1^{'}
-0.2196\psi_2^{'}-0.2984\psi_3^{'}
-0.2514\psi_4^{'} -0.1844\psi_5^{'}.
\eear
For the higher numbers of $N$ than $N=5$,
we found the similar convergent properties
in the expression of $\bar\psi_0$.
We also checked that the other orthogonal states,
$\bar\psi_n$, $(n\ge 1)$,
have the convergent series in their expansion coefficients
by raising $N$.
Using the expression of the orthogonal basis for a given $N$,
we constructed the truncated Schnabl's solution (\ref{tcSS2}).
We calculated the normalized tachyon potential $f(N)$
in Eq.~(\ref{ntp}) and found the same results in Table 1.
%%%%%%%%%%%%%%%%%%%%%%%%%%%%%%%%%%%%%%%%%%%
%%%%%%%%%%%%%%%%%%%%%%%%%%%%%%%%%%%%%%%%%%%
\begin{table}\label{converg}
\begin{center}\def\st{\vrule height 2.3ex width 0ex}
\begin{tabular}{|c||c|} \hline
$N$ & $\bar\lambda_n$
\st\\[0ex] \hline
$0$ & 0.31496
\st\\[0ex] \hline
$5$ &0.41439,\, 0.27719,\, 0.072777\, 0.018645,\, 0.00087062,\,
0.000027315
\st\\[0ex] \hline
$10$ & 0.40674,\, 0.29151,\, 0.085571,\,
0.054505,\, 0.0090014,\,0.0014360,\,
\st\\
&  0.000089022,\, 5.9890$\times 10^{-6}$,\,
9.6021$\times 10^{-8}$,\, 1.1696$\times 10^{-9}$,\,
$-3.8026\times 10^{-10}$
\st\\[0ex] \hline
 & 0.39960,\, 0.29724,\, 0.090905,\, 0.064312,\,
0.017602,\,0.0046710,\,
\st\\
$15$&  0.00055827,\, 0.000073178,\,  4.3180$\times 10^{-6}$,\,
 3.2883$\times 10^{-7}$,\,3.4230$\times 10^{-9}$,\,
\st\\
& 6.7699$\times 10^{-11}$,\,
5.7708$\times 10^{-13}$,\,
7.2692$\times 10^{-15}$,\, 2.9800$\times 10^{-17}$,\,
$-2.9633\times 10^{-9}$
\st\\[0ex] \hline
& 0.39775,\, 0.29756,\,
0.096908,\, 0.064655,\, 0.022717,\,0.0083524,\,
\st\\
$20$&0.0013873,\, 0.00025064,\,
0.000024051,\,
2.8430$\times 10^{-6}$,\,1.2781$\times 10^{-7}$,\,
\st \\
&
8.1196$\times 10^{-9}$,\,
8.3458$\times 10^{-11}$,\,
3.3754$\times 10^{-12}$,\,
7.0465$\times 10^{-14}$,\,
1.7739$\times 10^{-15}$,\,
\st \\
&2.5832$\times 10^{-17}$,\,
3.3640$\times 10^{-19}$,\,
2.3055$\times 10^{-21}$,\,
8.8003$\times 10^{-24}$,\,
$-3.3857\times 10^{-8}$
\st\\[0ex] \hline
&
0.39739,\, 0.29702,\, 0.099841,\,
0.065190,\, 0.024882,\,0.011712,\,
\st\\
& 0.0023989,\, 0.00053188,\,
0.000067479,\, 0.000010217,\,7.6109$\times 10^{-7}$,\,
\st\\
$25$&
8.0343$\times 10^{-8}$,\,
1.5590$\times 10^{-9}$,\,
9.2397$\times 10^{-11}$,\,
2.8421$\times 10^{-12}$,\,
1.3550$\times 10^{-13}$,\,
\st \\
&4.0610$\times 10^{-15}$,\,
1.3445$\times 10^{-16}$,\,
3.0310$\times 10^{-18}$,\,
6.6446$\times 10^{-20}$,\,
9.9971$\times 10^{-22}$,\,
\st \\
&1.3095$\times 10^{-23}$,\,
1.0964$\times 10^{-25}$,\,
6.5095$\times 10^{-28}$,\,
1.6845$\times 10^{-30}$,\,
$-5.5178\times 10^{-9}$
\st \\[0ex] \hline
\end{tabular}
\end{center}
\caption{{\small Eigenvalues of $\tilde Q_{N}$ for
various $N$. }}
\label{egnvl}
\end{table}
%%%%%%%%%%%%%%%%%%%%%%%%%%%%%%%%%%%%%%%%%%%
%%%%%%%%%%%%%%%%%%%%%%%%%%%%%%%%%%%%%%%%%%%

As we see in Table 2, the smallest eigenvalue for a given truncation
number is negative. The first negative one appears from
$N=9$ with magnitude $10^{-8}$. The second one
appear from $N=98$. In our numerical range (up to
$N=200$), there are only two negative eigenvalues.
The negative eigenvalue $\bar\lambda_-$
decreases with oscillating behaviors for small $N$ (up to about $N=50$),
and decreases very slowly for large $N$. $\bar\lambda_-$
almost stays around $10^{-8}$ in the range of our numerical experiments.
The properties of the second one are similar to
those of the first one.

%%%%%%%%%%%%%%%%%%%%%%%%%%%%%%%%%%%%%%%%%%%%%%%%%%%%%%%%%
%%%%%%%%%%%%%%%%%%%%%%%%%%%%%%%%%%%%%%%%%%%%%%%%%%%%%%%%%
\begin{table}\label{cvtst}
\begin{center}\def\st{\vrule height 2.7ex width 0ex}
\begin{tabular}{|c|c|c|c|c|c|} \hline
 & $N$=10 & $N$=20 & $N$=30 & $N$=40 & $N$=50
\st\\[0ex]
\hline
$\bar\lambda_0$
&0.4067444 & 0.3977491 & 0.3973819 & 0.3974832 & 0.3975486
\st\\[0ex]
\hline
$\bar\lambda_1$
&0.2915061 & 0.2975642 & 0.2965582 & 0.2960965 & 0.2959405
\st\\[0ex]
\hline
$\bar\lambda_2$
&0.08557135 & 0.09690845 & 0.1009738 & 0.1013045 & 0.1010974
\st\\[0ex]
\hline
$\bar\lambda_3$
&0.05450536 & 0.06465539 & 0.06625414 & 0.06825249 & 0.06937157
\st\\[0ex]
\hline
$\bar\lambda_4$
&0.009001405 & 0.02271652 & 0.02550463 & 0.02572638
& 0.02654655
\st\\[0ex]
\hline
$\bar\lambda_5$
&0.001435974 & 0.008352415 & 0.01446823 & 0.01789735
& 0.01898159
\st\\[0ex]
\hline\hline
& $N$=60 & $N$=70 & $N$=80 & $N$=90 & $N$=100
\st\\[0ex]
\hline
$\bar\lambda_0$
&0.3975807 & 0.3975964 & 0.3976044 & 0.3976086 & 0.3976110
\st\\[0ex]
\hline
$\bar\lambda_1$
&0.2958885 & 0.2958715 & 0.2958665 & 0.2958658 & 0.2958664
\st\\[0ex]
\hline
$\bar\lambda_2$
&0.1009304 & 0.1008443 & 0.1008076 & 0.1007954 & 0.1007940
\st\\[0ex]
\hline
$\bar\lambda_3$
&0.06986604 & 0.07005325 & 0.07010889 & 0.07011381 & 0.07010217
\st\\[0ex]
\hline
$\bar\lambda_4$
&0.02762730 & 0.02848060 & 0.02905543 & 0.02941796
& 0.02963724
\st\\[0ex]
\hline
$\bar\lambda_5$
&0.01910793 & 0.01913351 & 0.01924661 & 0.01943154
& 0.01964400
\st\\[0ex]
\hline
\end{tabular}
\end{center}
\caption{{\small Several biggest eigenvalues,
$\bar\lambda_0,\,\bar\lambda_1,\, \cdots \bar\lambda_5$,
for $N=10,\,20,\, \cdots 100$.}}
\label{cvg}
\end{table}
%%%%%%%%%%%%%%%%%%%%%%%%%%%%%%%%%%%%%%%%%%%%%%%%%%%%%%%%%%%%%%%%
%%%%%%%%%%%%%%%%%%%%%%%%%%%%%%%%%%%%%%%%%%%%%%%%%%%%%%%%%%%%%%%%

The negative mode corresponds to the smallest
eigenvalues for given $N$.
Here we are dealing with approximation of
infinite dimensional operator $\tilde Q$ with a sequence of finite
dimensional one, and the approximation works in such a way
that it matches approximately
with the biggest eigenvalues and corresponding eigenvectors.
To make sure this fact we investigate the validity of
the negative eigenmode concretely.
In order to figure out the negative mode
for $\tilde Q$ in terms of the behaviors of eigenvalue,
we need numerical data for very large number of $N$.
However, it is a difficult problem because of computation time
in the computer program. Instead of the eigenvalue, we investigated
the behavior of coefficients in the negative eigenmode
$\bar\psi_- = \sum_{n=0}^{N}U_{-n}\psi_n^{'}$ given in Eq.~(\ref{obs1}).
In the numerical work up to $N=200$,
we found that 6 coefficients for lowest states,
$U_{-i}$, ($i=0,\cdots, 5$), converge.
Several convergent coefficients are given in Table 4.
For the coefficients for the higher states, we could not find
convergent behaviors since they become irregular by raising $N$.
We fitted the convergent coefficients and take the limit $N$ goes to
infinity. We found the quantity
$\langle\bar\psi_{-}, \, \tilde Q,\bar\psi_{-}\rangle$ is positive
for the resulting coefficients in $N\to\infty$ limit.
This result implies that the negativity of $\langle\bar\psi_-,\,
\tilde Q\,\bar\psi_-\rangle$ comes from the contribution
of the higher states in $\bar\psi_-$, which have non-convergent
coefficients. From this result, we can see that
eigenvector $\bar\psi_-$ and its eigenvalue are not meaningful
in our approximation.\footnote{We are indebted
to Martin Schnabl on this point.}

Since we are considering spacetime independent
fluctuations around the vacuum,
the energy for the fluctuations is identified as
\bear\label{ED}
\Delta E = \frac12 \langle\tilde\Psi,\,
\tilde Q\tilde \Psi\rangle =
\frac12\sum_{m=0}^{N} \bar\lambda_m\bar c_m^2 =-\tilde S_0.
\eear
Here $\Delta E$ corresponds to the energy difference
from the vacuum energy.
In our numerical work, all fluctuations which have convergent
eigenvalues and eigenvectors with convergent coefficients
have positive contributions to $\Delta E$.
Therefore, our numerical result supports that the Schnabl's
vacuum solution is a minimum energy solution and stable for
off-shell fluctuations also.

\subsection{Potentials for various fields around the tachyon vacuum}
In the previous subsection, we calculated the quantity
$\langle\tilde\Psi,\,\tilde Q\tilde \Psi\rangle$ with spacetime
independent gauge fixed
fluctuation $\tilde\Psi$ on ${\cal H}^{{\rm vac}}_N$.
And we obtained the result (\ref{PQP3}) numerically.
Inserting the Eq.~(\ref{PQP3}) into the action (\ref{AC4})
which is defined around the tachyon vacuum, we obtain
\bear\label{AC5}
\tilde S_0(\bar c_m) = -\frac12 \sum_{m=0}^N \bar\lambda_m\bar c_m^2.
\eear
%%%%%%%%%%%%%%%%%%%%%%%%%%%%%%%%%%%%%%%%%%%%%%%%%%%%%%%%%%%%%%%
\begin{figure}\label{Fig2}
\centerline{\includegraphics[width=140mm]{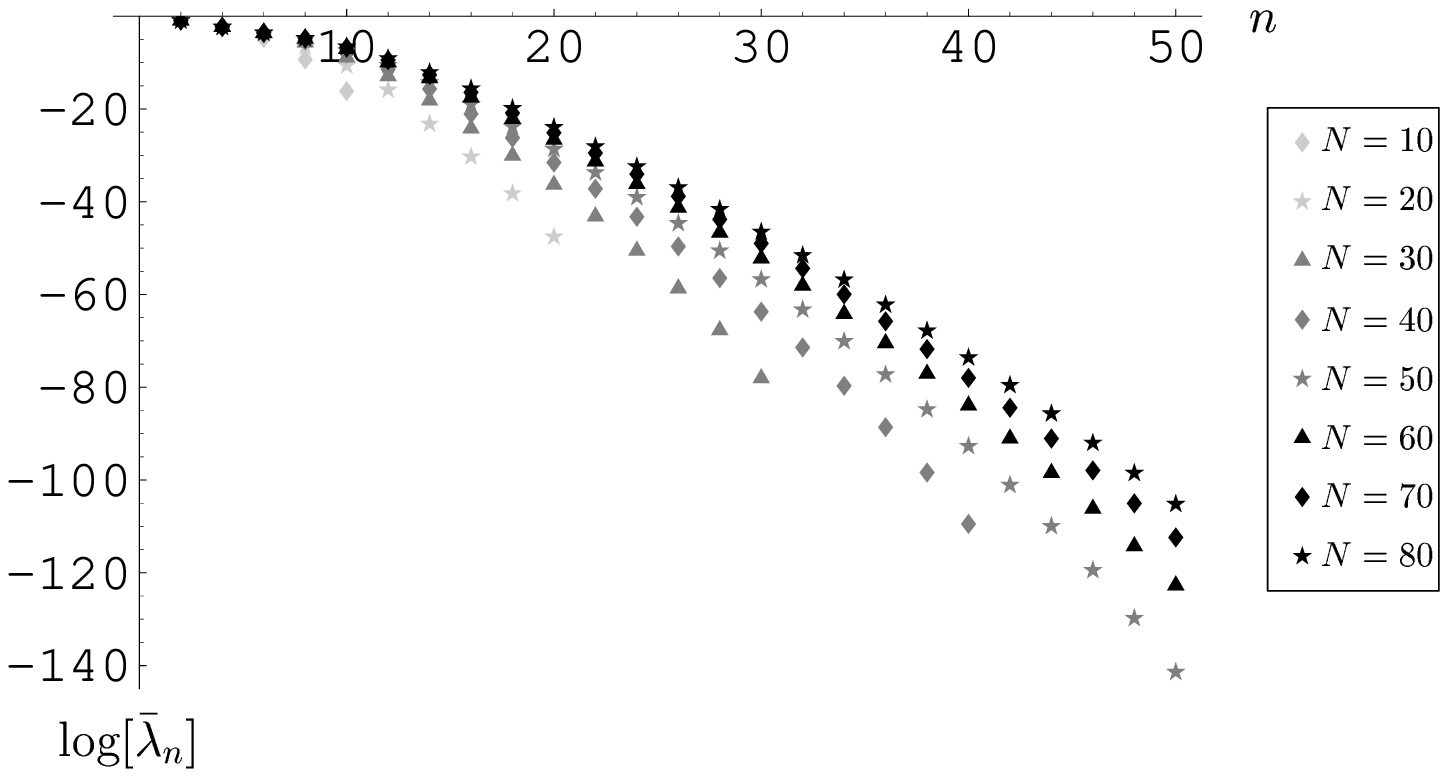}}
\caption{{\small Graphs of $\log\bar\lambda_n$ for $n=2,\, 4,\,
6,\, \cdots 50$ on the truncated subspace ${\cal H}_N^{{\rm vac}}$,
$(N=10,\, 20,\, \cdots 80)$.}}
\end{figure}
%%%%%%%%%%%%%%%%%%%%%%%%%%%%%%%%%%%%%%%%%%%%%%%%%%%%%%%%%%%%%%%%
%%%%%%%%%%%%%%%%%%%%%%%%%%%%%%%%%%%%%%%%%%%%%%%%%%%%%%%%%
%%%%%%%%%%%%%%%%%%%%%%%%%%%%%%%%%%%%%%%%%%%%%%%%%%%%%%%%%
\begin{table}\label{negvect}
\begin{center}\def\st{\vrule height 2.7ex width 0ex}
\begin{tabular}{|c|c|c|c|c|} \hline
 & $N$=100 & $N$=120 & $N$=140 & $N$=160
\st\\[0ex]
\hline
$U_{-0}$
&$0.0021283$ & $0.0023351$ &$0.0024864$ & $0.0025841$
\st\\[0ex]
\hline
$U_{-1}$
&$-0.0085477$ & $-0.0092470$ & $-0.0097477$ & $-0.010049$
\st\\[0ex]
\hline
$U_{-2}$
&$0.047870$ & $0.051303$ & $0.053727$ & $0.055103$
\st\\[0ex]
\hline
$U_{-3}$
&$-0.18174$ & $-0.19797$ & $-0.19904$ & $-0.20257$
\st\\[0ex]
\hline
\end{tabular}
\end{center}
\caption{{\small Coefficients for the several lowest states
in the negative eigenmode }}
\label{negvec}
\end{table}
%%%%%%%%%%%%%%%%%%%%%%%%%%%%%%%%%%%%%%%%%%%%%%%%%%%%%%%%%%%%%%%%
%%%%%%%%%%%%%%%%%%%%%%%%%%%%%%%%%%%%%%%%%%%%%%%%%%%%%%%%%%%%%%%%
In this expression, $\tilde S_0$ is defined to be
the action value divided by the spacetime volume factor
according to the convention of BPZ inner product in this paper.
Since we are considering spacetime independent fluctuations,
$\tilde S_0$ can be written as the potential density
around the tachyon vacuum,
\bear\label{POT}
V(\bar\phi_m) = - \tilde S(\bar\phi_m) =
\frac{1}{2}\sum_{m=0}^N\bar\lambda_m\bar\phi_m^2,
\eear
where we replace the arbitrary coefficients $\bar c_m$
with spacetime independent off-shell fields $\bar\phi_m$.
For several largest eigenvalues, for instance,
the potentials for independent fields are given by
\bear\label{POT2}
V(\bar\phi_m)=0.39761\,\bar\phi_0^2 +0.29587\,\bar\phi_1^2
+0.10082\,\bar\phi_2^2+ 0.070038\,\bar\phi_3^2
+0.029882\,\bar\phi_4^2+ \cdots,
\eear
where we used the data for $N=200$ case.
The explicit numbers of $\bar\lambda_m$ for
several truncation numbers $N$ were given in Table 2.
The eigenvalues $\bar\lambda_m$ with fixed $N$
exponentially decrease with small oscillating behaviors as
we increase $m$.
For example, in the case $N=200$, $\bar\lambda_m$ has the following
fitting curve,
\bear\label{FIT}
\bar\lambda_m \sim e^{-0.6102\, m}.
\eear

\section{Conclusion}

We have investigated
the behaviors of the quadratic fluctuations
around the tachyon vacuum on the truncated
subspace ${\cal H}_N^{{\rm vac}}$ numerically.
We showed that the truncated form of Schnabl's solution $\tilde \Psi_N$ is
well-behaved on ${\cal H}_N^{{\rm vac}}$
and has nice convergence property by raising $N$ and high accuracy
in BPZ inner product for large $N$.
The physics around the vacuum is governed by $\tilde S_0(\tilde\Psi)$
given in Eq.~(\ref{AC4}). In this paper we restricted our interest to the
spacetime independent quadratic fluctuation $\tilde \Psi$.
To calculate $\tilde S_0(\tilde\Psi)$ on
${\cal H}_N^{{\rm vac}}$, we constructed the orthogonal string state
$\bar\psi_m$, $(m=0,1,2,\cdots N)$, using the symmetric structure of
$\tilde Q$ and obtained corresponding eigenvalues $\bar\lambda_m$.

The eigenvalues $\bar\lambda_m$ have nice convergence properties
by raising $N$ also for small $m$.
As we increase the truncation number $N$, the number of meaningful
eigenvalues become large.
In our numerical results, most of eigenvalues are positive but
very small number of negative
eigenvalues appear. The first one with magnitude $\sim 10^{-8}$ appears
from $N=9$ and the magnitude of it very slowly grows according to
the truncation number $N$.
The second one appear from $N=98$ and has the same properties
with the first one.
As we argued in subsection 3.2, the negative modes are numerical
artifacts of our setting.
Thus all spacetime independent fluctuations around the vacuum
have positive contribution to energy in the range of our
numerical work.
This result supports that the Schnabl's vacuum solution is
stable and represents minimum energy solution
for off-shell fluctuations also.

Since we have taken into account the orthogonal basis states,
the corresponding fields for the states have no interactions with
other fields around the vacuum. Then the action $\tilde S_0(\tilde\Psi)$
on ${\cal H}_N^{{\rm vac}}$ with spacetime independent fluctuation
$\tilde\Psi$ corresponds to sum of quadratic forms of potentials
with coefficients $\bar\lambda_m$ for the fields as given in
Eq.~(\ref{POT}).
In canonical kinetic term with second order
derivatives in field theory,
there exist massive physical excitations
for harmonic  oscillator potential and
$\bar\lambda_m$ corresponds to ${\rm mass}^2$ of the field $\bar\phi_m$.
However, these phenomena do not happen since the absence of
physical state including tachyon state at the vacuum was proved
analytically~\cite{Ellwood:2006ba}. Thus the shapes of quadratic
potentials in our numerical results represent that
the kinetic term at the
tachyon vacuum has different form from the canonical
second order differential operator
and does not allow the physical excitations.

Extension of our work to the fluctuations
with nonvanishing momentum will be helpful
to figure out the role of kinetic
term around the vacuum and
to understand universal mechanism of vanishing of
physical excitations by comparing with
other theories, such as
boundary string field theory, $p$-adic
string theory, and DBI-type effective field
theory, etc.

\section*{Acknowledgements}
We are grateful to Chanju Kim, Yuji Okawa, and  Ho-Ung Yee for very
useful discussions. We also thank the referee for the helpful comments
for the improvement of  the manuscript.
This work was supported by the Science Research Center Program of
the Korea Science and Engineering Foundation through
the Center for Quantum Spacetime(CQUeST) of
Sogang University with grant number R11-2005-021. The work of OK was
partially supported by SFI Research Frontiers Programme.

\end{document}